# Localization dynamics in a binary two-dimensional cellular automaton: the Diffusion Rule

Genaro J. Martíne[1,2],z Andrew Adamatzky[2]
Harold V. McIntosh[3]

[1] Departamento de Posgrado, Escuela Superior de Cómputo, Instituto Politécnico Nacional, México.
E-mail: `genaro.martinez@uwe.ac.uk`
`http://uncomp.uwe.ac.uk/genaro/`
[2] Faculty of Computing, Engineering and Mathematical Sciences, University of the West of England, Bristol, United Kingdom.
E-mail: `andrew.adamatzky@uwe.ac.uk`
`http://uncomp.uwe.ac.uk/adamatzky/`
[3] Departamento de Aplicación de Microcomputadoras, Instituto de Ciencias, Universidad Autónoma de Puebla, Puebla, México.
E-mail: `mcintosh@servidor.unam.mx`
`http://delta.cs.cinvestav.mx/~mcintosh/`

**Abstract**

We study a two-dimensional cellular automaton (CA), called *Diffusion Rule* (DR), which exhibits diffusion-like dynamics of propagating patterns. In computational experiments we discover a wide range of mobile and stationary localizations (gliders, oscillators, glider guns, puffer trains, *etc*), analyze spatio-temporal dynamics of collisions between localizations, and discuss possible applications in unconventional computing.

---

# 1   Introduction

In our previous studies on minimal cellular automaton (CA) models of reaction-diffusion chemical system we constructed [4] a binary-cell-state eight-cell neighborhood 2D CA model of a quasi-chemical system with one substrate, state 0, and one reagent, state 1. In that model chemical reactions were represented by semi-totalistic transition rules. Every cell switches from state 0 to state 1 depending whether sum of neighbors in state 1 belongs to some specified interval or not. A cell remains in state 1 if sum of neighbors in state 1 belongs to another specified interval.

From 1296 cell-state transition rules, we selected a set of rules with complex behavior [4]. Amongst the complex rules, namely in $G$-class of morphological classification [4], we located so-called Diffusion Rule. CA governed by this rule often exhibits slowly non-uniformly growing patterns, resembling diffusive patterns in chemical systems with non-trivial coefficients of diffusion, or reaction-dependent diffusion coefficients, so the name of the rule.

The rule simulates sub-excitable [36] medium-like mode of perturbation propagation — cell in state 0 takes state 1 if there are exactly two neighbors in state 1, otherwise the cell remains in state 0, and, conditional inhibition — cell in state 1 remains in state 1 if there are exactly seven neighbors in state 1, otherwise the cell switches to state 1.

In present paper we are trying to answer the following questions. Is there a reaction-diffusion binary-state CA that express complex dynamic? Can we demonstrate that CA exhibits non-stationary growth of reaction-diffusion patterns? Do stationary or mobile generators of localizations, glider guns, exist in binary-state reaction-diffusion CA? Can the reaction-diffusion CA simulate an effective procedure and therefore be universal?

CA with space-time dynamics similar to that in spatially extended chemical systems are studied from early days of CA theory and applications [35], however most rules discovered so far lack minimality (some of the rules employ dozens of cell-states). Methods of selecting the rules also widely vary depending on theoretical frameworks, e.g. probabilistic spaces [20, 38] and genetic algorithms [14, 32]. Therefore, we envisage a strong need for a systematic analysis of propagating patterns like those observed in the Diffusion Rule. The propagating patterns are of upmost importance in modern computer science because such patterns play a vital role in developing novel and emerging computing paradigms and architectures, particularly collision-based computing [1, 3, 23].

We must mention that various authors have already obtained pioneering results in the studied rule. Magnier *et al.* discovered three primary gliders [33]. David Eppstein found four gliders known, already incorporated in our framework, and four new gliders which were novel for us (Fig. 4 (q) (t) and (u)).[1] The glider traveling along diagonals of the lattice (Fig. 4 (v)) was firstly recorded by Amling in 2002 (see Eppstein's web site). Finally, a glider gun and three puffer trains were discovered by Wótowicz.[2]

---

[1] See Eppstein's findings at `http://fano.ics.uci.edu/ca/rules/b2s7/`

[2] `http://www.mirwoj.opus.chelm.pl/ca/rules/life_2.gif`

Diffusion Rule CA is just one of many complex CA[3] exhibiting mobile localizations. Other famous examples include semi-totalistic rules as the Game of Life [17, 10], Brain's-brain and Critters rules [34], High Life [9], Life 1133 [22], Life Without Death [18], and Life variant $B35/S236$.[4] Other variant is with Larger than Life [16] and the Beehive and Spiral rules hexagonal CA [39, 5, 6], more recent candidates were proposed by George Maydwell with Hexagonal Life and Hexagonal Long Life rules.[5] Amongst 3D binary state CA supporting gliders Life 4555 and Life 5766 by Carter Bays [7, 8] are most widely known. In 1D there are Rule 110 [13, 37, 30, 26] and Rule 54 [11, 21, 25] CA, which support an impressive range of mobile localizations.

Our paper is structured as follows. In Sect. 2 we introduce basic concepts of CA model under investigation. Section 3 introduces results of statistical analysis of the Diffusion Rule using mean field theory. In Sect. 4 we present basic structures discovered in the Diffusion Rule CA. Section 5 compiles a catalogs of non-trivial interactions between mobile localizations, which could be used to designing basic elements of collision-based computers. In Sect. 6 we highlight our achievements in analysis of the Diffusion Rule and prospects for future studies.

## 2 Basic notations

We study family of 2D binary-state cellular automaton (CA) defined by tuple $\langle \mathbb{Z}^2, \Sigma, u, f \rangle$, where $\mathbb{Z}$ is the set of integers, every cell $x \in \mathbb{Z}^2$ has eight neighbors, orthogonal and diagonal (i.e. classical Moore's neighborhood) $u(x) = \{y \in \mathbb{Z} : x \neq y \text{ and } |x - y| \leq 1\}$, $\Sigma = \{0, 1\}$ is the set of *states*, and $f$ is a local transition function defined as follows:

$$x^{t+1} = f(u(x^t)) = \begin{cases} 1, & \text{if } (x^t = 0 \text{ and } \sigma_x^t \in [\theta_1, \theta_2]) \text{ or } (x^t = 1 \text{ and } \sigma_x^t \in [\delta_1, \delta_2]) \\ 0, & \text{otherwise} \end{cases}$$

(1)

where $\sigma_x^t = |\{y \in u(x) : y^t = 1\}|$, and $\theta_1, \theta_2, \delta_1, \delta_2$ are some fixed parameters such that $0 \leq \theta_1 \leq \theta_2 \leq 8$ and $0 \leq \delta_1 \leq \delta_2 \leq 8$.

We can write the rule as $R(\delta_1 \delta_2 \theta_1 \theta_2)$ or like $B\theta_1 \ldots \theta_2 / S\delta_1 \ldots \delta_2$ more traditional code. Also, the rule can be interpreted as a simple discrete model of a quasi-chemical system with substrate '0' and reagent '1', where $[\theta_1, \theta_2]$ is an interval of reaction, or association between substrate and reagent, and $[\delta_1, \delta_2]$ is an interval of dissociation. The family of rules includes Conway's Game Life, when intervals $[\delta_1, \delta_2]$ and $[\theta_1, \theta_2]$ are interpreted as intervals of survival and birth, respectively.

---

[3] http://uncomp.uwe.ac.uk/genaro/otherRules.html
[4] http://www.ics.uci.edu/~eppstein/ca/b35s236/
[5] Several interesting candidates in hexagonal representation are proposed by Andrew Wuensche using *DDLAB* at http://www.ddlab.com/ and by Maydwell using *SARCASim* at http://www.collidoscope.com/ca/

In our previous paper [4] we morphologically classified all 1296 rules, and studied how changes in parameters $R(\delta_1\delta_2\theta_1\theta_2)$ of cell-state transition rule influence space-time dynamics. For example, we discovered [4] a small subset of rules *Life* $2c22$,[6] $2 \leq c \leq 8$, which could be interpreted as quasi-chemical precipitating systems. For parameter set $[\theta_1, \theta_2]=[22]\,[1, 4]$, the system is transformed into $2^+$-medium, CA model of excitable system in sub-excitable mode.

We have found a cluster of semi-totalistic rules supporting structures of the Diffusion Rule. They are $B2/S2\ldots 8$ called *Life* $dc22$[7] where $d$ and $c$ take values between 2 and 8, and $d \leq c$. Therefore, we found that the rule $B2/S7$ or $R(7722)$ exhibits most reach dynamics of localized patterns amongst all the rules studied by us. Rules of the local transition are simple:

1. A cell in state 0 will take state 1 if it has exactly two neighbors in state 1, otherwise cell remains in state 0.

2. A cell in state 1 remains in state 1 if it has exactly seven neighbors in state 1, otherwise cell takes state 0.

## 3 Mean field approximation

Mean field theory is a proved technique for discovering statistical properties of CA without analyzing evolution spaces of individual rules [20, 28, 12]. The method assumes that elements of the set of states $\Sigma$ are independent, uncorrelated between each other in the rule's evolution space. Therefore we can study probabilities of states in neighborhood in terms of probability of a single state (the state in which the neighborhood evolves), thus probability of a neighborhood is the product of the probabilities of each cell in the neighborhood. Using this approach we can construct mean field polynomial for a semi-totalistic evolution rule [24] as follow:

$$p_{t+1} = \sum_{v=\delta_1}^{\delta_2} \binom{n-1}{v} p_t^{v+1} q_t^{n-v-1} + \sum_{v=\theta_1}^{\theta_2} \binom{n-1}{v} p_t^{v} q_t^{n-v} \qquad (2)$$

where $n$ represents the number of cells in neighborhood, $v$ indicates how often state 1 occurs in Moore's neighborhood, $n-v$ shows how often state 0 occurs in the neighborhood, $p_t$ is a probability of cell being in state 1, $q_t$ is a probability of cell being in state 0.

On the basis of outcomes of computational experiments we can suggest intervals of extreme densities of initial random configurations which leads to the emergence of localizations in the Diffusion Rule. In the lower limit best densities $d$ are $0.004 < d < 0.015$ and in the upper limit they are $0.992 < d < 0.997$ for the first 15–20 steps of evolution (Fig. 1). CA starting its evolution in random configuration with lower density of 1-states exhibit stationary or mobile self-localizations (like gliders or oscillators) at the beginning of the evolution, however in many cases collisions between mobile localizations leads to catastrophes, when 1-state patterns spread all over the lattice. Random initial configurations with higher (upper limit) density of 1-states produce either vanishing reactions between localizations or symmetrical growing patterns emerged as unions of two or more gliders.

---

[6] http://uncomp.uwe.ac.uk/genaro/Diffusion_Rule/life_2c22.html
[7] http://uncomp.uwe.ac.uk/genaro/Life_dc22.html

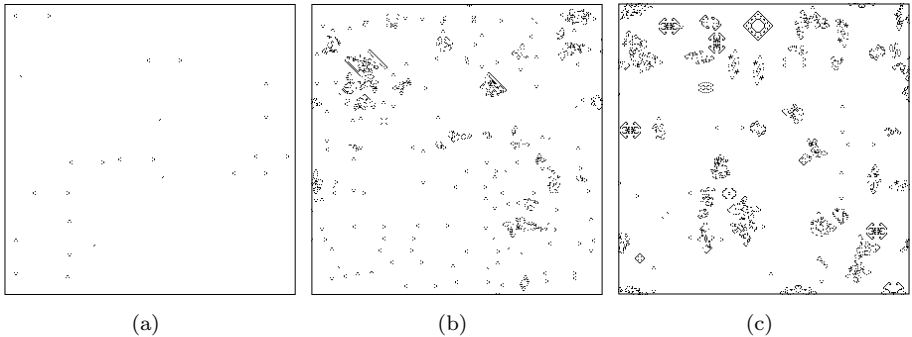

(a) (b) (c)

Figure 1: Three random initial densities for the Diffusion Rule: (a) 0.004, (b) 0.013 and (c) 0.995 respectively, on lattices of $200 \times 200$ to 18 generations.

Thus, the mean field polynomial for the Diffusion Rule is following:

$$p_{t+1} = 8p_t^8 q_t + 28p_t^2 q_t^7 \tag{3}$$

The fixed point is 0.236 that represents configurations with large density of 1-states emerging from any random initial condition (we should note that the fixed point for Conway's Game of Life is 0.37); this represents global density of 1-states necessary for evolution dynamics to stabilize. Also, we can see an unstable fixed point 0.05 (Fig. 2), that implies the existence of regions with unpredictable behavior or complex dynamic [28].

Thus, $p = 0$ is a super-stable point, although it is quite close to the unstable point. The super-stable is important, as it means a quiescent substrate, i.e., the state where live all the structures. Since mean field theory is just an initial approximation, it ought to be worthwhile to gather up some Monte Carlo approximations for a few generations just to see if the first estimate maintains itself, more or less.

We must also mention that the probability to find 'interesting' behavior is very low, about 0.05. Perhaps, this may be the reason why the Diffusion Rule was not studied before – when observing evolution from random configuration one more likely (with probability 0.3) to encounter a catastrophe (e.g. when placing three cells anywhere in Moore's neighborhood, just not in one line) then stable mobile localization.

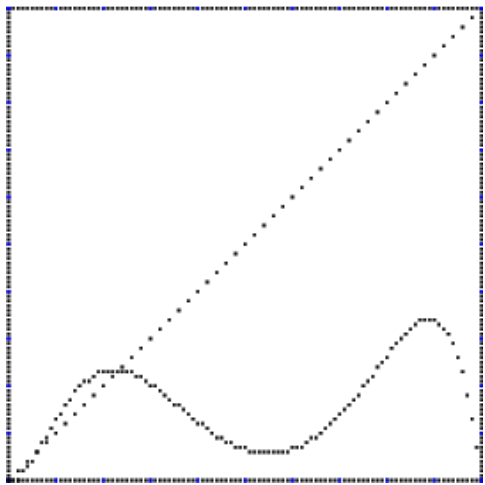

Figure 2: Diagram of mean field polynomial for the Diffusion Rule.

## 4 The Diffusion Rule Universe

In present section we uncover a range of basic structures, stationary and mobile localizations, generators of localizations and polymer-like structures formed of the mobile localizations.

### 4.1 Mobile self-localizations

In computational experiments with the Diffusion Rule CA we discovered 26 mobile self-localizations — gliders or particles — traveling orthogonally or diagonally in the lattice. Properties of the gliders, including volume, speed, direction of motion are listed in Table 1.

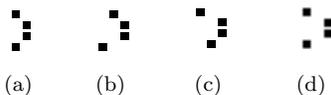

(a)     (b)     (c)     (d)

Figure 3: Configurations of minimal, or primary, gliders in the Diffusion Rule: (a) g1 glider, (b) g2 glider, (c) g3 glider and (d) g4 glider.

Configurations of minimal gliders and compound gliders are shown in Figs. 3 and 4, respectively. From Table 1, we can see that 96% of gliders move orthogonally, and that g23, g24, g25 and g26 are the largest gliders in the family of mobile localizations. First column in the table gives names of gliders. Second column in Table 1 represents glider's volume calculated as number of cells occupied by the glider. Third and fourth columns are translation and period.

Fifth column shows the speed=$c$/period, where $c$ is the maximum speed. Sixth column is the weight that represents the number of cells with state 1 within glider's volume. The last column indicates whether or not glider moves along columns and rows, or diagonals.

| glider | volume | translation | period | speed | weight | move |
|--------|--------|-------------|--------|-------|--------|------|
| g1  | 8   | 1 | 1 | $c/1$ | 4  | orthogonal |
| g2  | 12  | 1 | 1 | $c/1$ | 4  | orthogonal |
| g3  | 12  | 1 | 1 | $c/1$ | 4  | orthogonal |
| g4  | 12  | 1 | 1 | $c/1$ | 4  | orthogonal |
| g5  | 30  | 4 | 4 | $c/1$ | 7  | orthogonal |
| g6  | 30  | 4 | 4 | $c/1$ | 7  | orthogonal |
| g7  | 45  | 4 | 4 | $c/1$ | 14 | orthogonal |
| g8  | 45  | 4 | 4 | $c/1$ | 14 | orthogonal |
| g9  | 56  | 4 | 4 | $c/1$ | 14 | orthogonal |
| g10 | 56  | 4 | 4 | $c/1$ | 14 | orthogonal |
| g11 | 70  | 4 | 4 | $c/1$ | 24 | orthogonal |
| g12 | 72  | 4 | 4 | $c/1$ | 14 | orthogonal |
| g13 | 75  | 4 | 4 | $c/1$ | 18 | orthogonal |
| g14 | 84  | 4 | 4 | $c/1$ | 24 | orthogonal |
| g15 | 96  | 4 | 4 | $c/1$ | 18 | orthogonal |
| g16 | 96  | 4 | 4 | $c/1$ | 22 | orthogonal |
| g17 | 96  | 4 | 4 | $c/1$ | 26 | orthogonal |
| g18 | 96  | 4 | 4 | $c/1$ | 30 | orthogonal |
| g19 | 112 | 4 | 4 | $c/1$ | 26 | orthogonal |
| g20 | 126 | 4 | 4 | $c/1$ | 26 | orthogonal |
| g21 | 144 | 4 | 4 | $c/1$ | 26 | orthogonal |
| g22 | 144 | 4 | 4 | $c/1$ | 30 | orthogonal |
| g23 | 210 | 4 | 4 | $c/1$ | 38 | orthogonal |
| g24 | 338 | 2 | 4 | $c/2$ | 52 | orthogonal |
| g25 | 405 | 2 | 4 | $c/2$ | 79 | orthogonal |
| g26 | 576 | 2 | 8 | $c/4$ | 75 | diagonal |

Table 1: Properties of gliders in the Diffusion Rule CA.

There are two types of gliders — *primary* and *compound* [38, 26]: a primary glider can not be decomposed into smaller mobile localizations, a compound glider is made of at least two primary gliders.

## 4.2 Oscillators

The Diffusion Rule CA exhibits five types of stationary localizations known as oscillators. The two most common flip-flops and three blinker patterns are shown in Fig. 5.

Flip-flop configurations are o1 and o2 oscillators both structures flipping at 45°. Blinkers of period four are o3, o4 and o5 oscillators. Table 2 shows basic characteristics of each oscillator.

Some of the oscillators and their assembles act as eaters, i.e. stationary patterns that annihilate gliders colliding to them (see examples of collisions in Sect. 5.3).

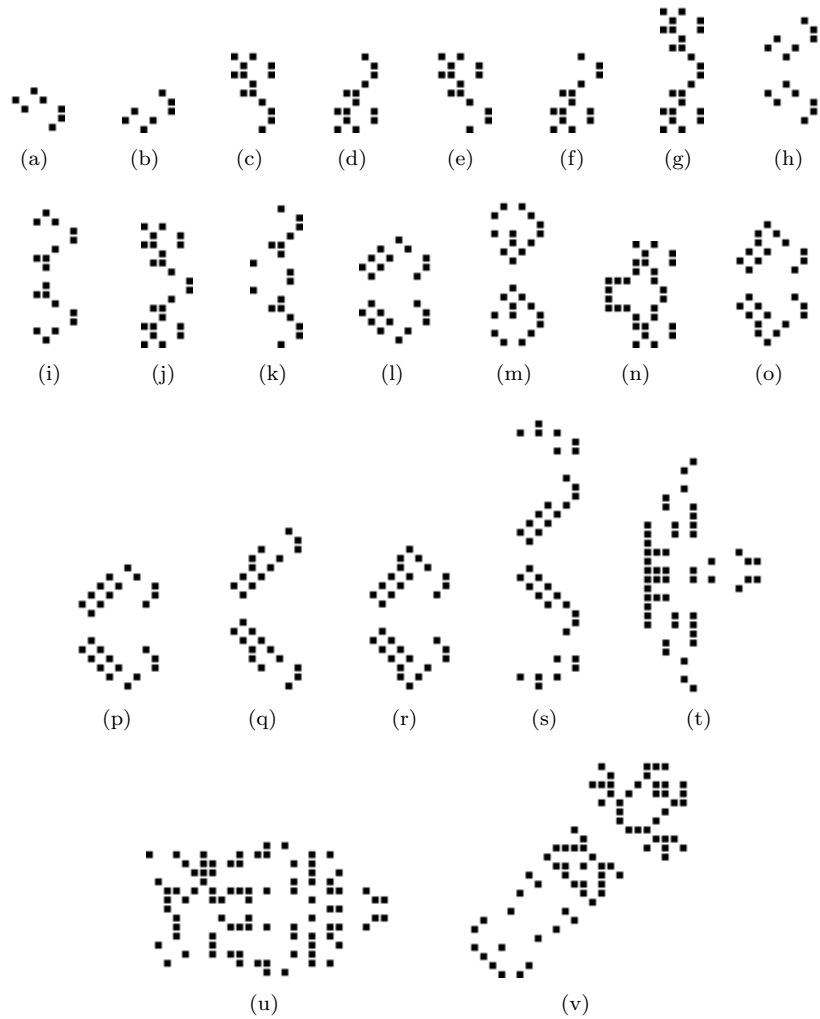

Figure 4: Twenty two compound gliders in the Diffusion Rule CA: (a) g5, (b) g6, (c) g7, (d) g8, (e) g9, (f) g10, (g) g11, (h) g12, (i) g13, (j) g14, (k) g15, (l) g16, (m) g17, (n) g18, (o) g19, (p) g20, (q) g21, (r) g22, (s) g23, (t) g24, (u) g25 and (v) g26 gliders, respectively.

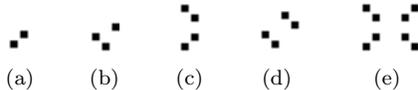

(a)  (b)  (c)  (d)  (e)

Figure 5: Five oscillators in the Diffusion Rule CA: (a) o1 and (b) o2 are flip-flops, (c) o3, (d) o4 and (e) o5 are blinkers, respectively.

| oscillator | volume | period | weight |
|---|---|---|---|
| o1 | 4 | 2 | 2 |
| o2 | 9 | 2 | 3 |
| o3 | 10 | 4 | 4 |
| o4 | 16 | 4 | 4 |
| o5 | 30 | 4 | 8 |

Table 2: Properties of oscillators in the Diffusion Rule CA.

### 4.3 Avalanches

Avalanches are novel structures, have not described before in related studies. They are assembles of adjacent gliders that cause explosive growth of rhomboid shaped patterns with deterministic edges and quasi-chaotic interior. Avalanches can be constructed from various compositions of adjacent gliders.

Figure 6 shows an avalanche produced in composition of two g1 gliders, adjacent at $90°$; this avalanche pattern grows diagonally inside the third quadrant. The minimal volume of an avalanche is $4 \times 4$ with eight cells in state 1.

One can use even number of gliders to construct symmetrical avalanches, two examples are shown in Fig. 7.

### 4.4 Puffer trains

A puffer train is a mobile localization which generate (leaves traces) of stationary localizations along its motion path. There are 16 known types of stable puffer trains (which produce oscillators) in the Diffusion Rule CA. Basic properties of puffer trains are shown in Table 3.

A particular case of puffer train is shown in Fig. 9. This puffer bears fragments of g4 glider. All configurations of the puffer are displayed in Fig. 9, and there we can see that configurations shown in Fig. 9 (c) (d) (e) and (f) can be interpreted as spaceships.

Moreover, the Diffusion Rule CA exhibits dozens of non-stable puffer trains. In the Fig. 10 we see five non-stable puffer trains, which produce asymmetrically growing, or quasi-chaotic, patterns. All discovered non-stable puffer trains have speed $1/c$ and period four.

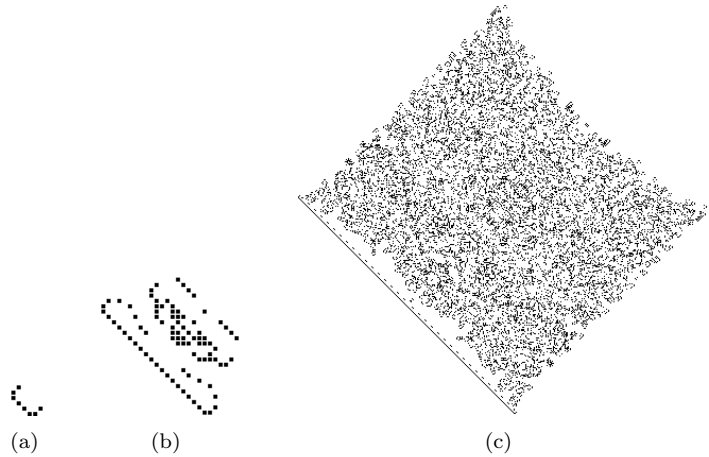

Figure 6: Two g1 gliders produce an avalanche pattern: (a) initial condition, (b) configuration after 15 steps of evolution and (c) configuration after 200 steps of evolution.

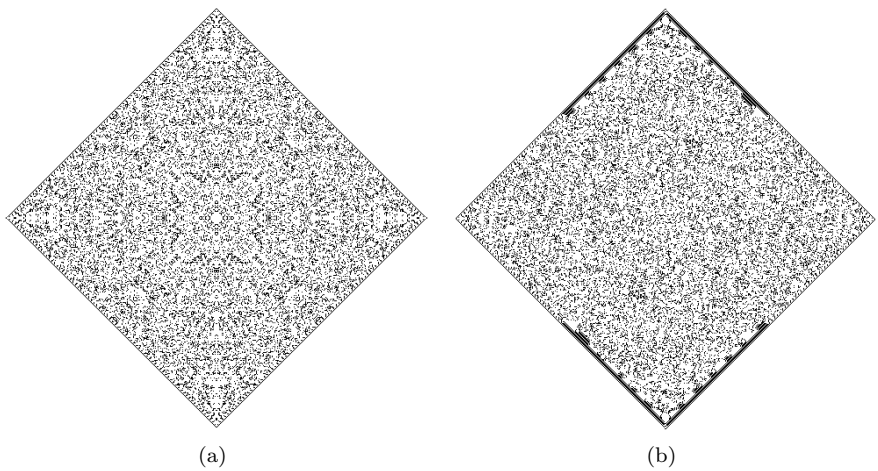

Figure 7: Construction of the symmetrical avalanches: (a) symmetrical growth initiated by two g4 gliders, configuration at 237th step of evolution, (b) avalanche pattern with non-trivial internal symmetries produced by assembly of two g2 glider and two g3 gliders, configuration at 237th step of evolution.

| puffer train | produce | volume | translation | period | speed | weight | move |
|:---:|:---:|:---:|:---:|:---:|:---:|:---:|:---:|
| p1 | o3 | 30 | 4 | 4 | $c/1$ | 7 | orthogonal |
| p2 | o1 | 35 | 4 | 4 | $c/1$ | 8 | orthogonal |
| p3 | o1 | 42 | 4 | 4 | $c/1$ | 9 | orthogonal |
| p4 | o1 | 42 | 4 | 4 | $c/1$ | 9 | orthogonal |
| p5 | o1 | 42 | 4 | 4 | $c/1$ | 10 | orthogonal |
| p6 | o1 | 54 | 4 | 4 | $c/1$ | 13 | orthogonal |
| p7 | o1 | 56 | 4 | 4 | $c/1$ | 9 | orthogonal |
| p8 | 2×o1 | 63 | 4 | 4 | $c/1$ | 14 | orthogonal |
| p9 | o1 | 63 | 4 | 4 | $c/1$ | 15 | orthogonal |
| p10 | $2n$×o1 | 65 | 4 | 4 | $c/1$ | 14 | orthogonal |
| p11 | 2×o1 | 80 | 4 | 4 | $c/1$ | 17 | orthogonal |
| p12 | o1 | 84 | 4 | 4 | $c/1$ | 11 | orthogonal |
| p13 | 2×o1 | 90 | 4 | 4 | $c/1$ | 11 | orthogonal |
| p14 | 2×o1 | 105 | 4 | 4 | $c/1$ | 15 | orthogonal |
| p15 | $2n$×o1 | 120 | 4 | 4 | $c/1$ | 28 | orthogonal |
| p16 | o1 $\vee$ $\epsilon$ | 242 | 4 | 4 | $c/1$ | 22 | orthogonal |

Table 3: Characteristics of puffer trains in the Diffusion Rule CA.

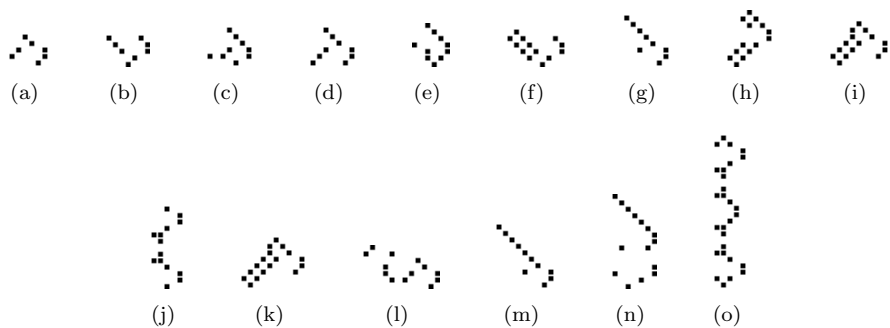

Figure 8: Fifteen puffer trains observed in evolution of the Diffusion Rule CA: (a) p1, (b) p2, (c) p3, (d) p4, (e) p5, (f) p6, (g) p7, (h) p8, (i) p9, (j) p10, (k) p11, (l) p12, (m) p13, (n) p14 and (o) p15 puffer trains, respectively.

### 4.5 Mobile glider guns

Glider guns are localized patterns that periodically lose their stability and give birth to traveling mobile localizations, gliders. In computational experiments we discovered twelve types of mobile glider guns in the Diffusion Rule CA.

Basic parameters of the twelve glider guns are shown in Table 4 and gun's configurations in Fig. 11. The most remarkable feature is that all primary gliders can be produced by glider guns. Some glider guns can generate two types of gliders at once, thus gun7 (Fig. 11f) generates g2 and g3 gliders at the same time, however both gliders travel coupled in pairs.

The generator gun12 can produce one or two g4 gliders, see examples in Fig. 12. This gun is also extendable, the extension is determined by the number of o1 oscillators (which should be more then two). Positions of oscillators

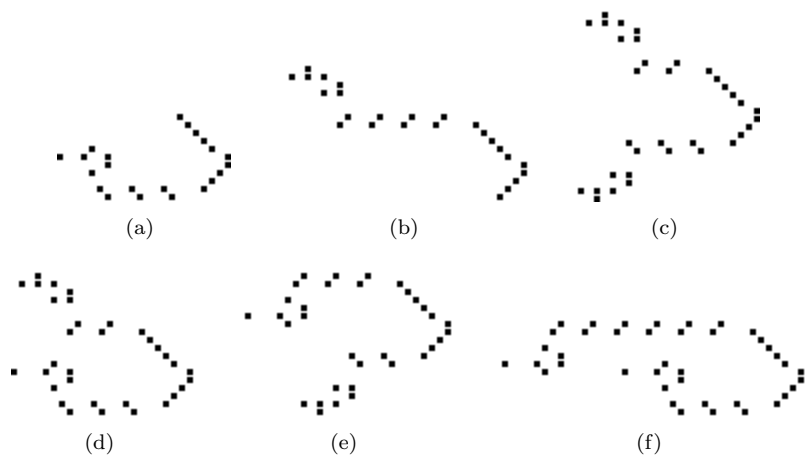

Figure 9: Special `p16` puffer train similar to spaceships.

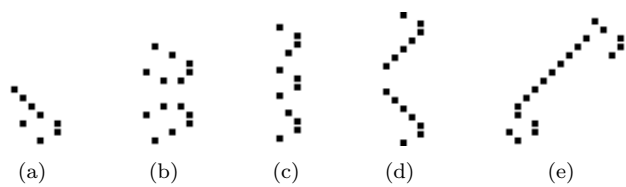

Figure 10: Some non-stable puffer trains.

| gun | produce | volume | translation | period | speed | weight | move |
| --- | --- | --- | --- | --- | --- | --- | --- |
| gun1 | g1 | 30 | 4 | 4 | $c/1$ | 7 | orthogonal |
| gun2 | g4 | 50 | 4 | 4 | $c/1$ | 9 | orthogonal |
| gun3 | g4 | 50 | 4 | 4 | $c/1$ | 9 | orthogonal |
| gun4 | g4 | 60 | 4 | 4 | $c/1$ | 9 | orthogonal |
| gun5 | g4 | 60 | 4 | 4 | $c/1$ | 9 | orthogonal |
| gun6 | g4 | 72 | 4 | 4 | $c/1$ | 15 | orthogonal |
| gun7 | g2 ∧ g3 | 72 | 4 | 4 | $c/1$ | 16 | orthogonal |
| gun8 | 2×g4 | 80 | 4 | 4 | $c/1$ | 14 | orthogonal |
| gun9 | 2×g4 | 90 | 4 | 4 | $c/1$ | 14 | orthogonal |
| gun10 | g1 ∧ g4 | 143 | 4 | 4 | $c/1$ | 15 | orthogonal |
| gun11 | 2×g1 | 154 | 4 | 4 | $c/1$ | 24 | orthogonal |
| gun12 | (2×g4) ∨ g4 | 176 | 4 | 4 | $c/1$ | 19 | orthogonal |

Table 4: Characteristics of glider guns in the Diffusion Rule CA. Second column of the table shows what type of glider each glider gun produces.

determine number of glider streams generated by gun12.

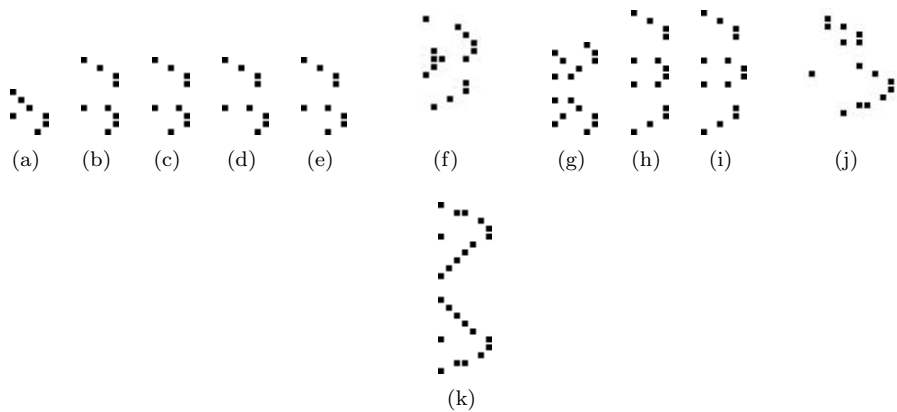

Figure 11: Configurations of glider guns in the Diffusion Rule CA.

So far we did not find glider guns which produce diagonally-moving streams of gliders, neither guns generating compound gliders or stationary guns.

## 4.6 Glider gun and puffer train

There is at least one special mobile structure that combines in itself properties of both glider gun and puffer train. Figure 13 shows glider gun producing g1 glider and o1 oscillator each 4th step of CA evolution. This puffer-gun moves orthogonally, has a volume of 70 cells and weights 12 units.

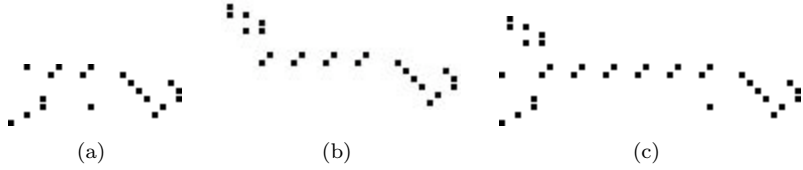

Figure 12: Extendable glider gun in the Diffusion Rule CA.

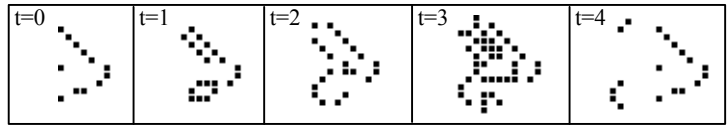

Figure 13: Configurations of puffer-gun in the Diffusion Rule CA.

### 4.7 Avalanche gun

Avalanche gun is another remarkable example of mobile generators (Fig. 14). The mobile gun produces an avalanche every 4th step of CA evolution.

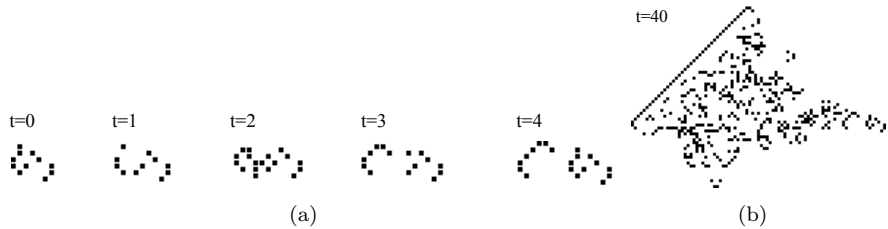

Figure 14: Configuration of an avalanche gun in the Diffusion Rule CA. (a) production of two g1 gliders to 90°, (b) development of quasi-chaotic reaction that destroys next avalanche conserving the avalanche gun.

However, life-time of the gun producing each avalanche is short: when avalanche produced by the gun it grows and then destroys the next avalanche produced.

## 5 Collisions between localized patterns

The Diffusion Rule CA combines high-degree unpredictability with enormously rich dynamics of collisions between mobile and stationary objects. This section studies the outcomes of the collision reactions.

## 5.1 Forming diffusing patterns by collisions

Gliders colliding in the Diffusion Rule CA can produce an explosively growing diffusion-like pattern, the diffusive patterns do usually have non-stationary boundaries and they exhibit quasi-chaotic internal dynamics.

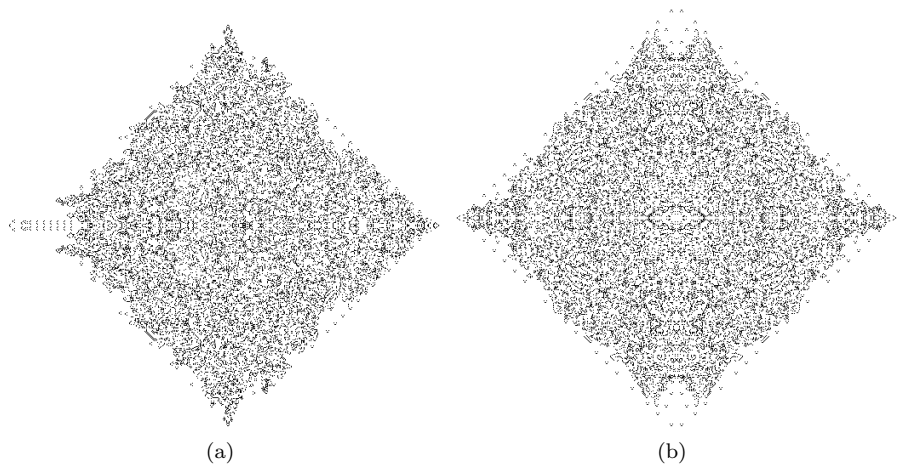

Figure 15: Examples of diffusion-like patterns produced in collision between gliders in the Diffusion Rule CA. (a) pattern produced at 260th step of evolution after collision between g4, g2 gliders and g3 glider, (b) pattern produced at 260th of evolution after collision between two g4 gliders and two g1 gliders.

Two most distinct examples are shown in Fig. 15. In first example (Fig. 15 (a)), g4 glider collides with g3 and g2 gliders, diffusion pattern produced is 'lead' by three gliders and puffer train. In second example (Fig. 15 (b)), two g4 gliders collides with two g1 gliders. The reaction produces multiple gliders, puffer trains, oscillators, and even vertical glider guns during their collision dynamics.

## 5.2 Reactions between propagating patterns

### 5.2.1 Soliton-like reaction

In certain initial conditions gliders collide similarly to solitons, namely they restore their structure and velocity vector after collisions. In Fig. 16 we can see snapshots of the head-on collision dynamics between two g4 gliders (begin at even distance before collision). When the gliders collide they temporarily lose their stability, produce varieties of transient structures. After few steps of evolution the gliders are restored and transient structures are annihilated.

At the moment, there is only a reaction soliton-like perhaps some others examples exist with more bigger gliders but initially have not more examples.

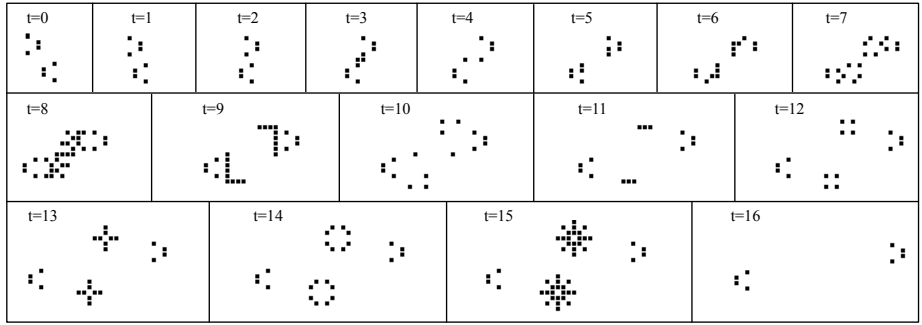

Figure 16: Soliton-like reaction between two g4 gliders.

#### 5.2.2 Eater reaction

Eaters are stationary localizations which destroy gliders colliding into them. The most simple eater is built with o1 oscillator. This eater destroys g1 gliders but is shifted four cells along the glider's initial direction of motion. Both g1 and g2 gliders are destroyed when they collide with the eater built of o3 oscillator. In Fig. 17(a) you can see an eater made of two o3 oscillators destroying g1 and g4 gliders. A 'universal' eater, which destroys all types of primary gliders is illustrated in Fig. 17(b).

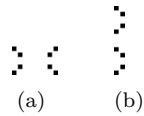

(a)    (b)

Figure 17: Examples of eater configurations in the Diffusion Rule CA.

Using eaters one can control glider streams emitted by glider guns, thus in Fig. 18 we can see how the eater eliminates g1 gliders produced by gun1. There is at least one mobile eater of gliders, this is g24 glider that eats g1 gliders.

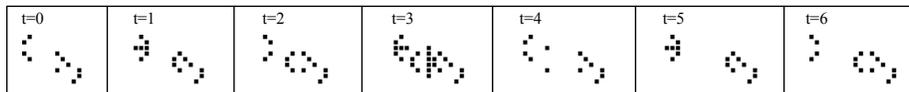

Figure 18: Eater destroys gliders emitted by glider gun.

#### 5.2.3 Delay reaction

The basic delay reaction can be implemented when g3 glider collides with g4 glider and changed to g4 glider in the result of collision, and the original g4

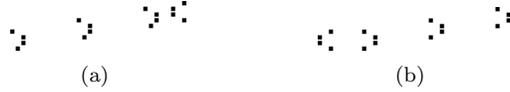

Figure 19: Delay reaction in the Diffusion Rule CA: (a) initial position of gliders before collision, (b) final result of delay operation, the glider traveling west delayed and translated southward.

glider delayed for two time steps but also translated southward one cell per collision (Fig. 19).

### 5.2.4 Multiplication and reduction reactions

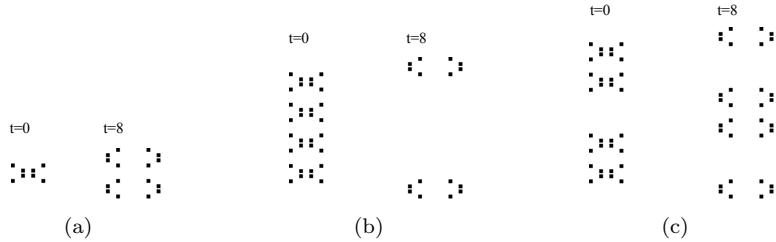

Figure 20: Multiplication of gliders: (a) multiplication $1 \times 1 \to 4$ reaction of gliders in binary collision, (b) reduction $4 \times 4 \to 4$ reaction in glider collision, (c) conservative collision reaction.

A typical reaction of glider multiplication is shown in Fig. 20 (a): two **g4** gliders are involved in head-on collision, with odd distance between glider heads before collision, four new **g4** gliders are produced in result of the collision. Reduction is implemented as multiplication of gliders, where gliders in the multiplied columns are in proximity of each other. As shown in Fig. 20 (b), when we collide two rows of **g4** gliders, four gliders in each row (eight gliders in total are involved in the collision), then just four new gliders are produced. Adjusting distance between gliders in colliding columns of gliders we can achieve almost any (but odd) result of multiplication (Fig. 20 (c)).

Using multiplication reactions we can also construct arbitrary packages of gliders. For example, to construct a stream of packages, six gliders per package, we collide stream of four-glider **g4** packages traveling west with a pair of **g4** gliders traveling east (Fig 21 (a)). Sequentially, all four-glider packages are transformed to six-glider packages (Fig. 21 (b)), the operation is halted by pair **g3** gliders.

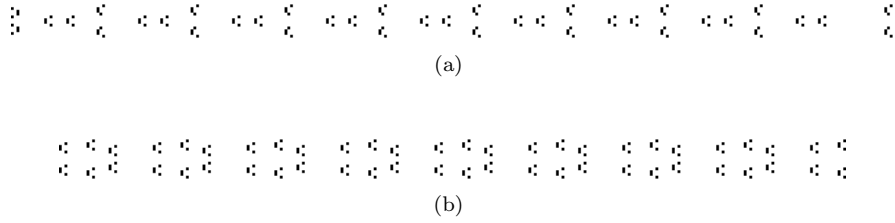

(a)

(b)

Figure 21: Constructing packages of g4 gliders by multiplication reaction. (a) initial configuration, (b) final configuration, recorded after 186th steps of evolution.

### 5.2.5 Reflection reaction

We discovered nine types of reflection-type collisions between stationary and mobile self-localizations. Let us discuss some examples shown in Fig. 22. When two g1 gliders collide with each other (head-on collision, even distance, with slight shift between gliders along south-north axis) two g4 gliders are generated; these g4 gliders move in the direction perpendicular to original trajectories of colliding gliders (Fig. 22 (a)). Glider colliding with o1 oscillator is reflected at the angle 90°, as shown for collision of g2, g4 and g3 gliders with o1 oscillator (Fig. 22 (b) (c) (d)).

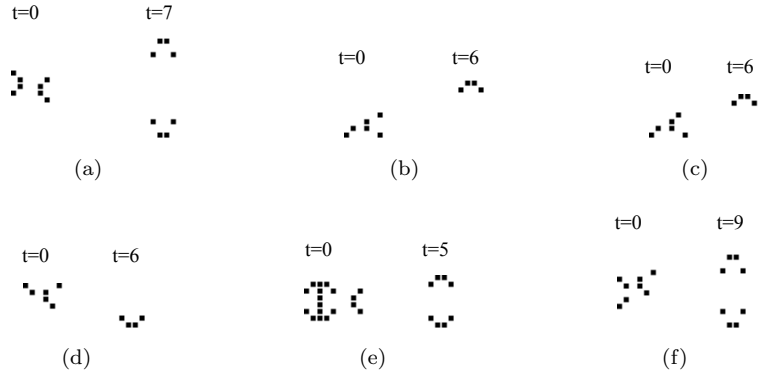

Figure 22: Collisions leading to reflections: (a) two g1 gliders collide with each other, (b) g4 glider collides with o1 oscillator, (c) g2 glider collides with o1 oscillator, (d) g3 glider collides with o1 oscillator, (e) g1 glider collides with o5 oscillator, (f) g3 glider collides with o3 oscillator

Gliders can also be derived in addition of reflections, when colliding with stationary localizations. Thus, when a g1 glider collides with a o5 oscillator, two g1 gliders (going in opposite directions to each other) are generated and follow trajectories perpendicular to original trajectory of collided g1 gliders (Fig. 22 (e)).[8] Similarly, when g3 glider collides with o3 oscillator, two g4 gliders are produced (Fig. 22 (f)).

### 5.2.6 Annihilation reaction

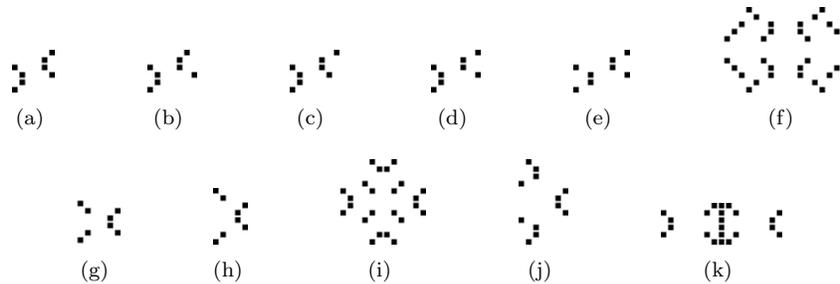

Figure 23: Initial positions of colliding gliders and oscillators leading to annihilation reaction. (a)–(f) binary collisions, (g)–(k) multiple collisions.

Significant amount of collisions between localizations in the Diffusion Rule CA leads to annihilation of colliding patterns. Few examples of initial configurations of colliding objects (leading to annihilation) are shown in Fig. 23, for binary collisions between gliders (Fig. 23 (a)–(f)) and multiple collisions between gliders and oscillators (Fig. 23 (g)–(k)).

## 5.3 Computation in the Diffusion Rule

Basic operations necessary to implement a functionally complete set of logical gates can be derived from collision dynamics presented in Fig. 22. Following paradigms of collision-based computing [2] we encode logical TRUTH by presence of a glider or an oscillator, while absence of mobile or stationary objects corresponds to logical FALSITY.

Namely, Fig. 22 (b) (c) (d) demonstrate that stationary localizations, oscillators, can play a role of mirrors thus deflecting gliders from their original trajectory. The mirrors can be used to route signals. Signals can be deleted, erased by placing eaters along trajectories of gliders, representing the signals. Signals can be also delayed in collisions with other gliders or oscillators.

Collisions used to construct FANOUT gate are shown in Fig. 22 (e) (f), a glider collides to stationary localization, and two new gliders are produced in result of the collision.

Dynamics displayed in Fig. 22 (a) shows a typical collision gate, where inputs $x$ and $y$ are represented by trajectories of gliders traveling East and West, respectively. While trajectories of new gliders, traveling South and North, encode value of $x$ AND $y$. Constant TRUTH is made up of ceaseless stream of gliders, generated by glider guns.

---

[8]This reaction can synchronize multiple collisions as you can see in our example FANOUT.rle available from http://uncomp.uwe.ac.uk/genaro/Diffusion_Rule/diffusionLife.html

### 5.3.1 Constructing a memory device

A memory in the Diffusion Rule CA can be constructed using basic interactions between g1 gliders and o1 oscillator, as illustrated in Fig. 24. A bit of information is represented in the memory unit by shaded $2 \times 2$ cells square in Fig. 24. The bit can be read by sending a g1 glider to the memory unit (top configurations in Fig. 24, glider travels East). When g1 glider collides with oscillator o1 oscillator, both glider and oscillator are annihilated (and one new o1 oscillator is formed at some distance from the memory unit). Then the bit is erased as a result of the reading operation. To write down the bit one can send another g1 glider (in Fig. 24 this 'writing' glider travels West) toward now empty memory unit and associated o1 oscillator. Both g1 glider and associated oscillator are destroyed, however o1 oscillator is restored in memory unit (shaded region in Fig. 24), i.e., we write a bit again.

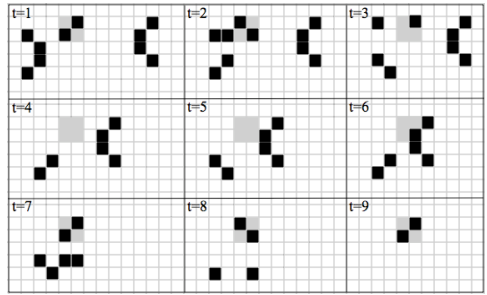

Figure 24: Constructing a memory device in the Diffusion Rule CA. Snapshots of the configurations. Time increases from right to left and from top to bottom. The domain where the bit of information (o1 oscillator) is written to is represented by a shaded zone.

### 5.3.2 Asynchronous XNOR and XOR gate

Exploiting some features of the interaction between g1 glider and o1 oscillator (Fig. 24) we can implement an *a*synchronous device which calculates XNOR and XOR operation at once (Fig. 25). Such a gate is designed by a scheme similar to that outlined in [6]. Oscillator in position shaded by gray in Fig. 24 represents logical value TRUE and absence of the oscillator — value FALSE of logical operation XNOR, exclusive NOR operation. An auxiliary oscillator (generated when oscillator in shade position is annihilated) represents results of XOR operation, exclusive OR operation.

Assuming that signals on inputs $x$ and $y$ can be generated at any time (no synchronization) except not at the same time, we obtain the following dynamics

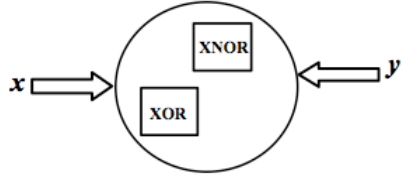

Figure 25: Scheme of XNOR and XOR gate.

of the device (Fig. 25):

| $x$ | $y$ | XNOR | XOR |
|---|---|---|---|
| 0 | 0 | o1 | 0 |
| 0 | g1 | 0 | o1 |
| g1 | 0 | 0 | o1 |
| g1 | g1 | o1 | 0 |

where `o1` and `g1` stay for oscillator and glider, 0 means absence of the objects.

## 6 Discussion

Findings discussed in the paper are based on computational experiments with the Diffusion Rule CA and, particularly, exhaustive search of mobile and stationary localizations emerged in spatio-temporal dynamics of the automaton.

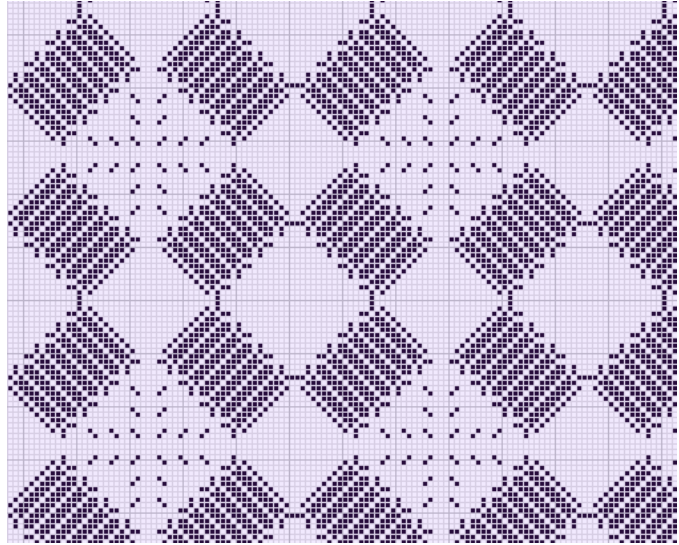

Figure 26: Simulating a luminescence pattern with the Diffusion Rule.

Amongst known 2D CA supporting localizations, the Diffusion Rule CA is the minimal model because cell-state transitions depend not on intervals of 'cell sensitivity' but on singletons, i.e. transition $0 \to 1$ occurs if there is exactly two neighbors in state 1, and transition $1 \to 1$ if there exactly seven neighbors in state 1. Moreover, we are not aware of any other CA which exhibits so large variety of mobile localizations (gliders) and high diversity of outcomes of collisions between mobile and/or stationary localizations. Despite of trying to undertake exhaustive study of localization dynamics we nevertheless missed several important points that could form objectives of future studies.

A simple but interesting physical simulation was made setting a reaction like luminescence. Using packages of both diagonal lines of 50 cells everyone. The luminescence phenomenon was obtained over the evolution in densities of cells in small intervals as illustrated the Fig. 26. The final state is dominated by blinkers or oscillators. This simulations are developed by the group of researchers iGEM-México at the MIT.[9]

It is not a trivial problem to find large stable patterns constructed with big complex structures. Cell colonies damaged by a virus is one of such configurations (Fig. 27).

Diffusion Rule can simulate the evolution of a cell like the elemental cellular automata (ECA) Rule 18 or Rule 90. For example, we take a diagonal line with 503 cells in the initial condition. During the evolution original line is multiplied in lines less and less small producing oscillators. Finally, the evolution space is dominated by oscillators that represent exactly a cell alive in 1D case. However, there is generated a second evolution of the same type as illustrated in Fig. 28. In this example, it took 512 steps of evolution to reach the configuration of 7,596 living cells.

The existence of stable configurations seems difficult to find in a rule which is generally chaotically producing super-nova explosions. Nevertheless, we have an example where four glider guns are synchronized to annihilate the gliders. In this case, two `g1` gliders and two `g4` gliders come into quadruple collision shown in Fig. 29.

We envisage that important open problems to be solved include implementation of quasi-chemical reactions between gliders, studies of grammars derived and implementation of a full effective decision procedures based on glider collisions. It will be also worth to demonstrate intrinsic universality and self-reproduction. Another project would be to use de Bruijn diagrams [29] to check if there are any still undiscovered gliders with velocity one or still life configurations, and besides to use algorithms specialized in automatic search for complicated or big gliders [15], oscillators, glider guns or more configurations. Also we are planning to make an exploration of the cluster of semi-totalistic rules originated by the Diffusion Rule.[10] Finally, the last but not least open problem is to decide if all types of gliders can be constructed in collision with other gliders, a closure property with respect to set of gliders.

---

[9]http://www.fenomec.unam.mx/pablo/igem/
[10]http://uncomp.uwe.ac.uk/genaro/Life_dc22.html

# Acknowledgements

We thank Cris Moore, David Hillman, Paul Callahan and Markus Redeker for their valuable comments. We also grateful to Dave Greene for fruitful discussions about complex constructions in the Diffusion Rule. Also we thanks to anonymous values commentaries from JCA board referees. First author still thanks support of EPSRC (grant EP/D066174/1).

# References

[1] Adamatzky, A. (2001) *Computing in Nonlinear Media and Automata Collectives*, Institute of Physics Publishing, Bristol and Philadelphia.

[2] Adamatzky, A. (Ed.) (2003) *Collision-Based Computing*, Springer.

[3] Adamatzky, A., De Lacy Costello B., & Asai T. (2005) *Reaction-Diffusion Computers*, Elsevier.

[4] Adamatzky, A., Martínez, G.J., & Seck Tuoh Mora, J.C. (2006) Phenomenology of reaction-diffusion binary-state cellular automata, *Int. J. Bifurcation and Chaos* **16 (10)** 1–21.

[5] Adamatzky, A., Wuensche, A. & De Lacy Costello, B. (2005) Glider-based computing in reaction diffusion hexagonal cellular automata, *Chaos, Solitons & Fractals*.

[6] Adamatzky, A. & Wuensche, A. (2006) Computing in spiral rule reaction-diffusion cellular automaton, *Complex Systems* **16 (4)**.

[7] Bays, C. (1987) Candidates for the game of life in three dimensions, *Complex Systems* **1** 373–400.

[8] Bays, C. (1991) New game of three-dimensional life, *Complex Systems* **5** 15–18.

[9] Bell, D.I. (1994) High Life — An interesting variant of life, `http://www.tip.net.au/~dbell/`.

[10] Berlekamp, E.R., Conway, J.H., & Guy, R.K. (1982) *Winning Ways for your Mathematical Plays*, Academic Press, (vol. 2, chapter 25).

[11] Boccara, N., Nasser, J., & Roger, M. (1991) Particle like structures and their interactions in spatio-temporal patterns generated by one-dimensional deterministic cellular automaton rules, *Physical Review A* **44 (2)** 866–875.

[12] Chaté, H. & Manneville, P. (1991) Evidence of collective behavior in cellular automata, *Europhysics Letters* **14** 409–413.

[13] Cook, M. (2004) Universality in elementary cellular automata, *Complex Systems*, **15 (1)** 1–40.

[14] Das, R., Mitchell, M. & Crutchfield, J.P. (1994) A genetic algorithm discovers particle-based computation in cellular automata, *Lecture Notes in Computer Science* **866** 344–353.

[15] Eppstein, D. (2002) Searching for spaceships, *MSRI Publications* **42** 433–452.

[16] Evans, K.M. (2003) Replicators and larger-than-life examples, (in ref. [19]).

[17] Gardner, M. (1970) Mathematical Games — The fantastic combinations of John H. Conway's new solitaire game Life, *Scientific American* **223** 120–123.

[18] Griffeath, D. & Moore, C. (1996) Life Without Death is P-complete, *Complex Systems* **10** 437–447.

[19] Griffeath, D. & Moore, C. (Eds.) (2003) *New constructions in cellular automata*, (Santa Fe Institute Studies on the Sciences of Complexity) Oxford University Press.

[20] Gutowitz, H.A. & Victor, J.D. (1987) Local structure theory in more that one dimension, *Complex Systems* **1** 57–68.

[21] Hanson, J.E. & Crutchfield, J.P. (1997) Computational mechanics of cellular automata: an example, *Physics D* **103 (1-4)** 169–189.

[22] Heudin, J.-K. (1996) A new candidate rule for the game of two-dimensional life, *Complex Systems* **10** 367–381.

[23] Jakubowski, M.H., Steiglitz, K., & Squier, R. (2001) Computing with Solitons: A Review and Prospectus, *Multiple-Valued Logic*, Special Issue on Collision-Based Computing **6 (5-6)**.

[24] Martínez, G.J. (2000) Teoría del Campo Promedio en Autómatas Celulares Similares a The Game of Life, Tesis de Maestría, CINVESTAV-IPN, México.

[25] Martínez, G.J., Adamatzky, A. & McIntosh, H.V. (2006) Phenomenology of glider collisions in cellular automaton Rule 54 and associated logical gates, *Chaos, Fractals and Solitons* **28** 100–111.

[26] Martínez, G.J., McIntosh, H.V., & Seck Tuoh Mora, J.C. (2006) Gliders in Rule 110, *Int. J. Unconventional Computing* **2 (1)** 1–49.

[27] McIntosh, H.V. (1988) Life's Still Lifes, http://delta.cs.cinvestav.mx/~mcintosh

[28] McIntosh, H.V. (1990) Wolfram's Class IV and a Good Life, *Physica D* **45** 105–121.

[29] McIntosh, H.V. (1994) Phoenix, http://delta.cs.cinvestav.mx/~mcintosh/oldweb/pautomata.html

[30] McIntosh, H.V. (1999) Rule 110 as it relates to the presence of gliders, http://delta.cs.cinvestav.mx/~mcintosh/oldweb/pautomata.html

[31] Minsky, M. (1967) *Computation: Finite and Infinite Machines*, Prentice Hall.

[32] Mitchell, M. (2001) Life and evolution in computers, *History and Philosophy of the Life Sciences* **23** 361–383.

[33] Magnier, M., Lattaud, C., & Heudin, J.-K. (1997) Complexity Classes in the Two-dimensional Life Cellular Automata Subspace, *Complex Systems* **11 (6)** 419–436.

[34] Tommaso, T. & Norman, M. (1987) *Cellular Automata Machines*, The MIT Press, Cambridge, Massachusetts.

[35] von Neumann, J. (1966) *Theory of Self-reproducing Automata* (edited and completed by A. W. Burks), University of Illinois Press, Urbana and London.

[36] Sediña-Nadal, I., Mihaliuk, E., Wang, J., Pérez-Muñuzuri, V., & Showalter, K. (2001) Wave propagation in subexcitable media with periodically modulated excitability, *Phys. Rev. Lett.* **86** 1646–49.

[37] Wolfram, S. (2002) *A New Kind of Science*, Wolfram Media, Inc., Champaign, Illinois.

[38] Wuensche, A. (1999) Classifying cellular automata automatically, *Complexity* **4 (3)** 47–66.

[39] Wuensche, A. (2004) Self-reproduction by glider collisions: the beehive rule, *Alife9 Proceedings*, 286–291, MIT Press.

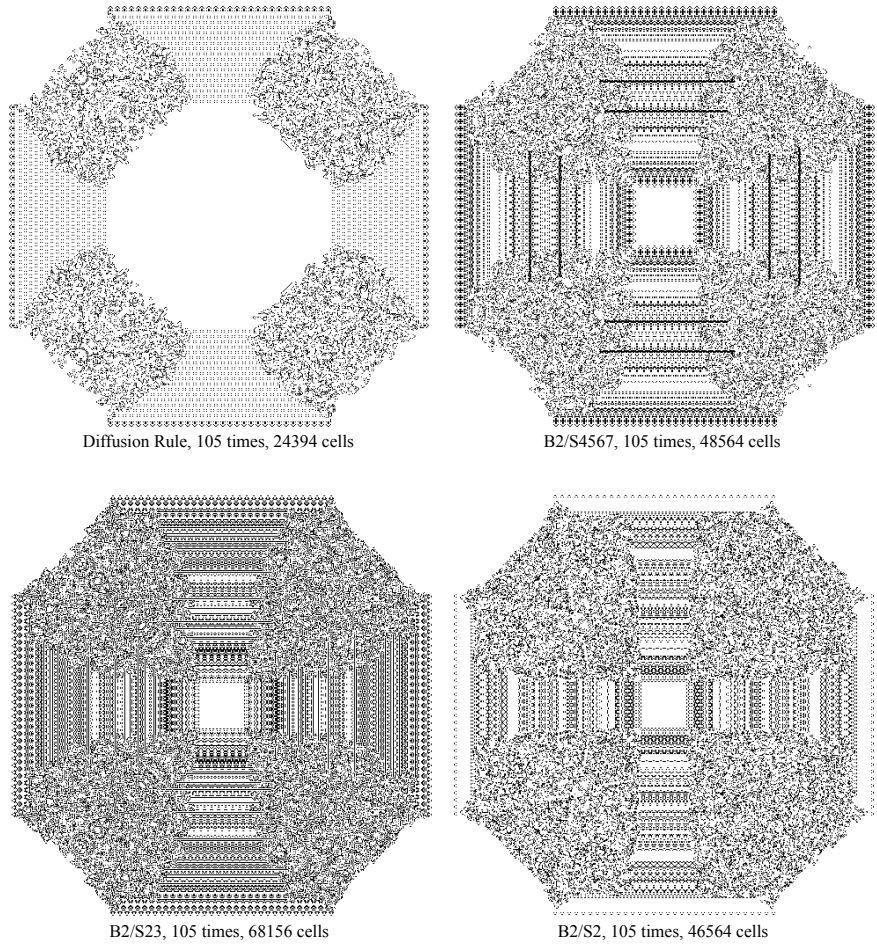

Figure 27: Virus propagation in the Diffusion Rule and three mutations in the same initial condition. Evolution rules and data are given below snapshots.

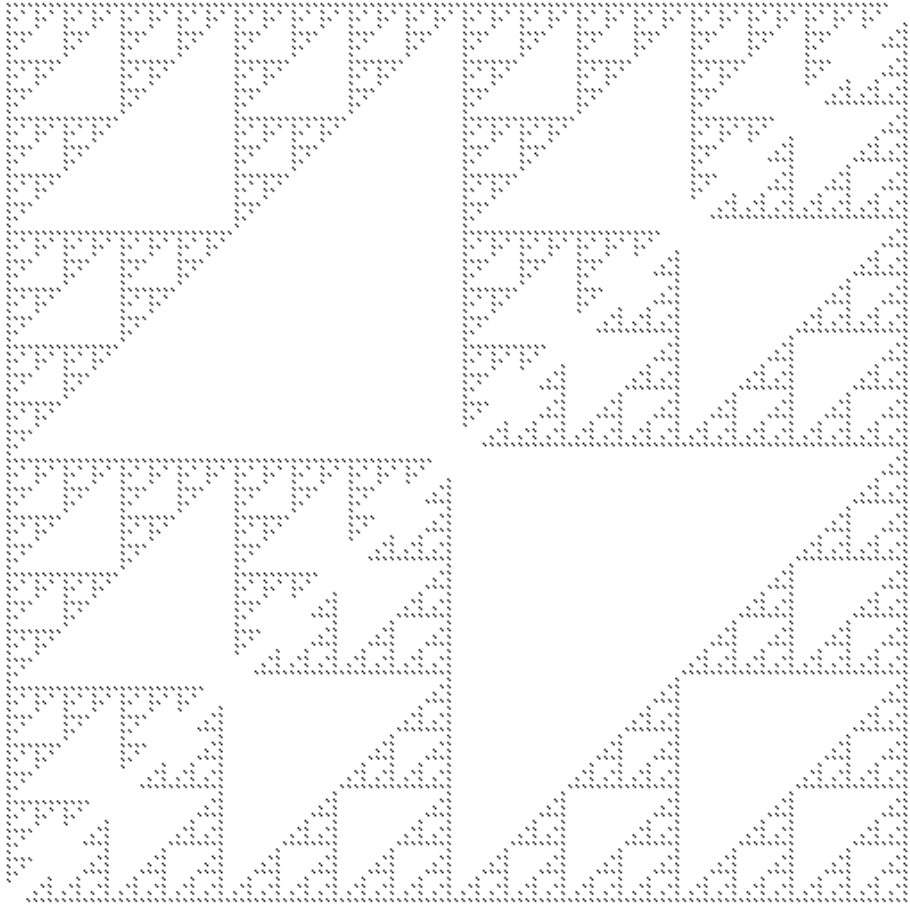

Figure 28: Simulating the evolution of a cell of the ECA Rule 18 or Rule 90 with the Diffusion Rule.

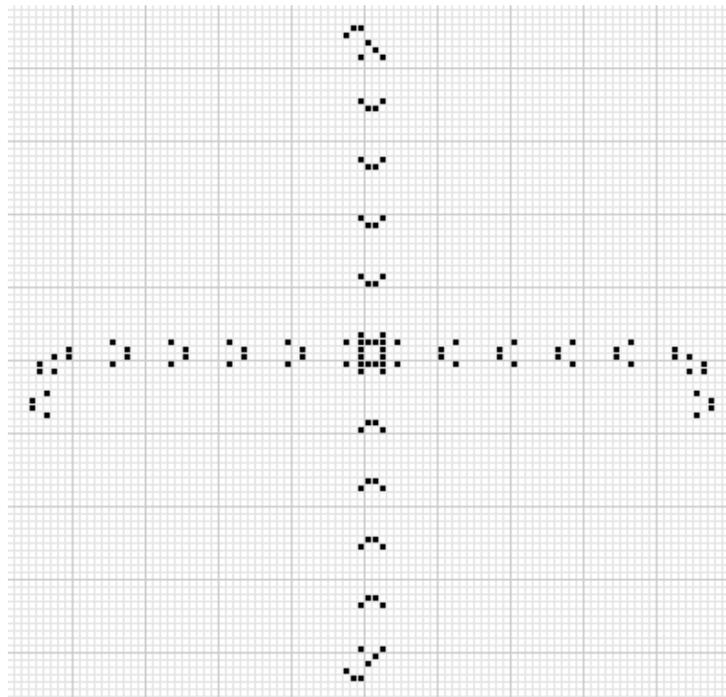

Figure 29: Simultaneous annihilation across of four particles produced by four glider guns in the Diffusion Rule. The evolution shows a global configuration in 36 steps with 118 live cells.